\documentclass[letterpaper, 10 pt, conference]{ieeeconf}  

\IEEEoverridecommandlockouts                              

\usepackage{graphics} 
\usepackage{amsmath} 
\usepackage{amssymb}  

\usepackage{flushend} 

\usepackage{subfigure}
\usepackage{svg}

\title{\LARGE \bf
Second-Order Coverage Control for Multi-Agent UAV Photogrammetry
}

\author{Samuel Mallick$^{1}$, Airlie Chapman$^{1}$, and Eric Schoof$^{2}$
\thanks{$^{1}$Samuel Mallick and Airlie Chapman are with the Department of Mechatronic Engineering,
        The University of Melbourne, 3052 Parkville, Australia
        {\tt\small smallick@student.unimelb.edu.au}, {\tt\small airlie.chapman@unimelb.edu.au}.}%
\thanks{$^{2}$Eric Schoof is with the Department of Electrical and Electronic Engineering, The University of Melbourne, 3052 Parkville, Australia {\tt\small eschoof@unimelb.edu.au}.}%
}

\begin{document}

\maketitle
\thispagestyle{empty}
\pagestyle{empty}

\begin{abstract}

Unmanned Aerial Vehicles equipped with cameras can be used to automate image capture for generating 3D models via photogrammetry. Current methods rely on a single vehicle to capture images sequentially, or use pre-planned and heuristic imaging configurations.
We seek to provide a multi-agent control approach to capturing the images required to 3D map a region. A photogrammetry cost function is formulated that captures the importance of sharing feature-dense areas across multiple images for successful photogrammetry reconstruction. A distributed second-order coverage controller is used to minimise this cost and move agents to an imaging configuration.
This approach prioritises high quality images that are simultaneously captured, leading to efficient and scalable 3D mapping of a region.
We demonstrate our approach with a hardware experiment, generating and comparing 3D reconstructions from image sets captured using our approach to those captured using traditional methods.

\end{abstract}

\section{INTRODUCTION}\label{intro}

Photogrammetry is the process of extracting 3D models from sets of images.
It is applied to generate high fidelity 3D models used in 3D mapping \cite{kalacska2021comparing}, surveying \cite{gomez2016uav} and other applications \cite{themistocleous2015methodology} \cite{bonora2021photogrammetry}.
Photogrammetry tools have become increasingly mature and accessible \cite{Westoby2012} \cite{MVS}, with free software packages now allowing amateur users to create high quality 3D models \cite{Alice} \cite{wu2011visualsfm}. 
In photogrammetry, images of a region of interest are captured from different camera positions.
A 3D point cloud representing points in the region is then reconstructed by triangulating matched feature points extracted from the images. 
The quality of the 3D model depends heavily on the input images.
Multiple images must overlap the same features for them to be captured in the reconstruction.
Insufficient feature overlap between images can result in images not being used in the reconstruction or causing the reconstruction to fail.
Additionally, the compute time for the reconstruction increases with the number of images.

As low-cost unmanned aerial vehicles (UAVs) equipped with cameras have become prevalent, UAVs, or agents, have been recognised as tools for capturing photogrammetric images \cite{nex2011uav}.
Additionally, multi-agent UAV networks have become increasingly popular in research and application \cite{pantelimon2019survey}. These networks can be applied to photogrammetry to parallelize image acquisition and reduce capture time \cite{Feichtner2018}.
Delegating image capture to agents, or groups of agents, requires particular attention to the agents' flight path or way-points to ensure a suitable image set is acquired.

\textbf{Literature Review.}
A single-agent approach is to pre-plan a flight path over an area of interest, typically with a simple uniform line or grid configuration of waypoints. An agent using an onboard camera then sequentially captures the image set from these waypoints.
Eisenbeiss \cite{eisenbeiss2008autonomous} used a UAV to capture images for 3D mapping. For flight planning they determined the plane of the region of interest and then covered this region uniformly with waypoints that had consistent image overlap between neighbouring image locations.
Remondino \textit{et al.} \cite{nex2011uav} describe UAV flight planning in reference to several real world archaeological mapping tasks in a survey. Waypoints are calculated to fix the longitudinal and transveral overlap between images and to uniformly cover the region.
Approaches that cover a region uniformly ignore variations in the region of interest and can require high quantities of images, some of which may be redundant, to guarantee sufficient overlap in the image set. 

Another single-agent approach is to use information generated from intermediate 3D reconstructions to inform future locations for image capture. These fall under the \emph{Next Best View} (NBV) literature. 
Huang \textit{et al.} \cite{huang2018active} present an automated image capturing process for 3D reconstruction using NBV planning. Their approach chooses a NBV to increase a completeness metric, formulated to evaluate the coverage of a 3D model.
Hepp \textit{et al.} \cite{hepp2018plan3d} demonstrate an end-to-end system for automated UAV mapping and 3D reconstruction. 
They model the region as a discretized occupancy grid and choose consecutive next viewpoints to maximise the information gained while considering collision-avoidance trajectory constraints. 
Peralta \textit{et al.} \cite{peralta2020next} give a reinforcement learning approach to generate next best image locations. Their algorithm learns the criteria for selecting imaging locations and results in 3D reconstructions that are more complete than those generated from uniformly spaced configurations. 
A limitation of NBV approaches is that intermediate reconstructions require time and compute during image capture. The approaches can also require large numbers of images and high computation for the final reconstruction.

 Various approaches have been explored to employ a network of camera-equipped agents for photogrammetry. Auer \textit{et al.} \cite{Feichtner2018} describes a hardware implementation of UAV networks to photogrammetry. They give three benefits of a network approach: the use of low cost off-the-shelf UAVs, reduced operation time and, as a consequence of the second point, less variation in environmental conditions during mapping. However they rely on pre-planned flight paths with the region either being partitioned between the UAVs or the UAVs moving together in formation. 
 Lundberg \textit{et al.} \cite{lundberg2018visualsfm} have demonstrated a UAV network 3D modelling framework, 
 focusing on elimination of the complex movement patterns required for single UAV modelling. Their framework requires a preliminary 3D reconstruction, which relies on a highly-capable and well-equipped `primary' UAV. 
 Chen \textit{et al.} \cite{chen2021multi} gives a UAV network approach to mapping environments using a clustered-collaborative-operation algorithm for scheduling tasks between the agents. However the algorithm relies on voxel reconstructions in real-time, requiring significant computation.

\textbf{Contribution.} This paper presents a novel multi-agent control strategy for positioning agents to capture quality images for 3D reconstruction using a network of UAVs.
This approach uses a fixed number of high quality images that are captured simultaneously, allowing a region to be reconstructed rapidly with a predictable capture and compute time.
Inspired by the coverage control literature, a photogrammetry cost function is formulated to quantify the performance of the network configuration for generating photogrammetry images.
This cost function incorporates a feature-density distribution over the region of interest that models the concentration of features to be extracted from images, and a sensor model that represents the overlap of features between UAV images.

A second-order coverage controller is used to control the UAVs. This controller is known to minimise an auxiliary cost, which in turn is shown to bound the photogrammetry cost.
Using this controller the agents converge to a configuration of camera locations that results in a high quality reconstruction of the region.
We demonstrate the convergence of agents to a locally optimal configuration and give experimental results comparing 3D reconstructions using images captured from the resultant configurations against traditional uniform configurations.

The key benefits of this method are as follows.
First, by modelling the environment with a feature density and maximising feature overlap, areas of high information are prioritised. Consequently high quality reconstructions that capture the significant elements of a region can be generated from fewer images.
Second, this approach uses a fixed number of images, optimizing the image set within capture time and compute constraints.
Finally, the proposed controller is absent of a centralised planner and can be distributed across agents reducing communication costs, and improving scalability and robustness to agent failure \cite{cortes2017coordinated}. 

This paper is structured as follows. Section \ref{formulation} gives some preliminary definitions and notation, relevant photogrammetry theory, and a formal problem statement. Section \ref{coverage_control} formulates a coverage cost function for photogrammetry and gives the distributed agent controllers to solve the problem. Section \ref{experiments} present simulations and physical experiments validating the approach. Finally, Section \ref{conclusions} provides concluding remarks and identifies directions for future work.

\section{PROBLEM FORMULATION} \label{formulation}
This section provides preliminary notation used throughout the paper, a problem statement and some of the basic theory of photogrammetry.

\subsection{Notation} \label{prelims}
The 2-norm of a vector $z$ is denoted as $\|z\|$ and the infinity-norm as $\|z\|_\infty$.
We consider a closed convex environment $Q \subset \mathbb{R}^2$. The \textit{diameter of $Q$} is defined as $\mbox{diam}Q = \max_{q,p \in Q}\|q-p\|$.
Let the positions of $n$ agents be $P = \{p_1, p_2, ..., p_n\}$ where $p_i \in Q$ for $i \in \{1, 2,...,n\}$. We define a set of ordered index pairs $C = \{(i,j) | i,j \in \{1, 2, ..., n\}, i < j\}$.
 Pairs of agent positions are denoted as $\Gamma_{ij} = \{p_i, p_j\}$ for each $(i,j) \in C$.

\subsection{Problem Description} \label{problem_spec}
We consider the task of creating a 3D model of the region $Q$ using photogrammetry. 
A multi-agent system of $n$ camera equipped UAV agents move to a desired imaging configuration. They then capture $n$ images simultaneously which are used in the photogrammetry reconstruction. 
Agents operate in a horizontal plane at a fixed height above the ground. It is assumed there are no obstacles to avoid and that agents can localise themselves in the world frame. 
It is further assumed that inter-agent collision avoidance is handled by a secondary controller, modifying control inputs for safety, and negligibly effecting the nominal control input.
The 2D position $p_i$ of each agent $i$, in the plane, is governed by single integrator dynamics
\begin{equation*}
    \dot{p}_i = u_i
\end{equation*}
where $u_i \in \mathbb{R}^2$ is a control input for agent $i$.
The camera of each agent faces downward with a field-of-view (FOV) area of radius $r$.

\textbf{Problem.} We seek agent controllers $u_i$ for $i = \{1,...,n\}$ that drive the agents, from arbitrary initial positions, to a configuration $P = \{p_1, p_2, ..., p_n\}$ where the captured images
result in a high quality photogrammetry reconstruction.
 
 \subsection{Photogrammetry Foundations}\label{formulation_pg}

In this section we will briefly review photogrammetry theory relevant to the design of a photogrammetry cost function in Section \ref{coverage_control}. 
For a full treatment of the theory, see \cite{Schindler2014MathematicalFO} and \cite {luhmann2019close}.

\subsubsection{World Point Reconstruction}
Photogrammetry constructs a point cloud model by using triangulation to reconstruct points captured in images.
A necessary condition for a given point in the world to be reconstructed is that it is captured and identified in \textit{two or more images}. 
These points in images are referred to as \textit{features}, and are identified using feature extraction algorithms, e.g., SIFT \cite{lowe2004distinctive}.
The need for matching features between images leads to a requirement of high feature overlap between images in a photogrammetry image set. This will be the foundation of the photogrammetry cost function formulated in Section \ref{coverage_control}.

\subsubsection{Basic Pipeline}

There are a several key stages in a photogrammetry pipeline to go from a set of images to a 3D point cloud. 
First a feature extraction step extracts all features from the images. Then a feature and image matching step identifies matching features and groups images that share features. At this stage images with insufficient feature overlap are discarded. Finally an iterative optimization is used to determine the location of matched features as 3D world points in a point cloud model.

\subsubsection{Photogrammetry Performance Measures} \label{formulation_measure}
In the 3D modelling literature there are several measures against which reconstructions are measured. 
Accuracy and precision of reconstructed 3D world points are used as measures when locations of world points are know \textit{a priori} \cite{fraser1984network} \cite{hoppe2012photogrammetric} \cite{olague2002automated}.
Point-cloud density of the 3D model, a measure of how many reconstructed 3D world points per unit of area or volume, is another measure against which 3D reconstructions can be evaluated \cite{rebolj2017point} \cite{dai2013comparison}. 
Completeness measures that evaluate how much of a physical model has been reconstructed are seen in the modelling of complex geometrical objects as they assess coverage of the object as well as holes and hidden areas \cite{huang2018active} \cite{wu2014quality}. 

We consider a fixed region to be mapped with a fixed number of images, and wish to maximise the information extracted from a photogrammetric reconstruction. 
Intuitively, more distinct 3D world points encodes more information of the modelled region. 
For this paper we will use the number of points in a 3D point cloud reconstruction as a performance measure for the photogrammetry reconstructions. 

\section{SECOND-ORDER COVERAGE PHOTOGRAMMETRY} \label{coverage_control}
To evaluate the performance of an agent configuration $P$, one option is agents capture images, create a 3D reconstruction and evaluate the point count.
This takes significant time and processing power for every evaluation, and does not inform a next, improved, configuration.
Instead we will formulate a cost function that can evaluate a configuration without a 3D reconstruction, by quantifying the overlap of feature-dense areas, and use this cost to design controllers for the agents.

\subsection{Coverage Cost Function for Photogrammetry}
To find such a cost function we turn to the coverage control literature.
Coverage control is a mature body of multi-agent literature addressing location optimisation for agents to effectively cover a region with their sensors \cite{cortes2004coverage} \cite{jiang2015higher}.
Each point in the region is assigned to one or more agents, as their responsibility to `sense' or `cover'.
The collections of points assigned to the same agents form cells for those agents, with the collection of cells forming a partition of the region.
A partition of the region $Q$ is a set of $m$ sub-regions $W = \{W_1, W_2, ..., W_m\}$ with disjoint interiors, $W_i \cap W_j = \emptyset$ for all $i \neq j$, whose union is $Q$, i.e., $\bigcup\limits_{i = 1}^{m} W_i = Q$.
When two agents are assigned to sense each point in the region $Q$, it is referred to as \textit{second-order coverage control}. The points assigned to agents $i$ and $j$ form the cell $W_{\Gamma_{ij}}$ and the set of all cells form a partition of $Q$; $W = \{W_{\Gamma_{12}}, W_{\Gamma_{13}},...,W_{\Gamma_{(n-1)(n)}}\}$.
The general form of the cost function for a set of positions $P$ and a partition $W$ is
\begin{equation*}
    H_f(P, W) = \sum_{\forall W_{\Gamma_{ij}} \in W} \int_{W_{\Gamma_{ij}}} f(\|q-p_i\|, \|q-p_j\|)\phi(q)dq
\end{equation*}
where $H_f$ decreases as the coverage performance increases.
The function $\phi: Q \to \mathbb{R}$ is a positive-valued function representing the relative importance of covering points in $Q$.
The sensor model $f: \mathbb{R}^2 \to \mathbb{R}$ describes the degradation in sensing performance as the distance to the sensing agents increases.
The subscript on $H_f$ emphasises that the cost is parameterised by the sensor model $f$.

We provide an interpretation of $\phi$ and $f$ that makes $H_f$ representative of the photogrammetry performance for an agent configuration $P$ and a partition $W$.
In photogrammetry the information to extract from the environment is image features that can be reconstructed as 3D points.
The density $\phi(q)$ describes an estimate of the relative number of features at point $q$ that can be extracted from an image.
The value of $\phi$ will be high in areas where there are many distinctive features to capture.
This feature density can be provided \textit{a priori} or estimated dynamically as the agents move to their final configuration \cite{schwager2009decentralized}, for example by applying a feature extraction algorithm (e.g., SIFT) to single images of an area.
The photogrammetry sensor model is given as 
\begin{equation}
\label{eq:photogrammetrySensorFunction}
    h(x, y) = \begin{cases}
    g(x,y) & \|\begin{bmatrix}x & y  \end{bmatrix}\|_\infty \leq r \\
    2(\mbox{diam}Q)^2 & \mbox{otherwise}
    \end{cases}
\end{equation}
where $g(x,y) = x^2 + y^2$, $x(p_i) = \|q - p_i\|$, $y(p_j) = \|q - p_j\|$, and $p_i, p_j \in Q$.
The value $2(\mbox{diam}Q)^2$ is chosen as the maximum value of $g(x,y)$ when $p_i, p_j \in Q$.
The sensor model captures the need for features to be present in two images to be reconstructed in the 3D model, as all points not overlapped by the FOV of both agents, i.e., when $\max({\|q-p_i\|, \|q-p_j\|}) > r$ for $q \in Q$, are penalised by a large constant penalty.
Points that are overlapped by the FOV of both agents are penalised quadratically with distance from the agents through the function $g(x,y)$.
This is inspired by the polynomial model for image distortion where
distortion increases with distance from the center of the image \cite{bukhari2013automatic}.
The goal of controlling the agent network for photogrammetry is captured in minimising this cost via the agent's configuration $P$ and the partitioning of the space between agents $W$.

\subsection{The Optimal Partition}
A second-order Voronoi partition of $Q$ with respect to $P$ is denoted as $V(P)^\textbf{2} = \{V_{\Gamma_{12}}, V_{\Gamma_{13}}, ..., V_{\Gamma_{(n-1)(n)}}\}$ where
 \begin{equation*}
     V_{\Gamma_{ij}} = \{q \in Q | \forall v \in \Gamma_{ij}, \forall w \in P\setminus\Gamma_{ij}, \|q - v \| \leq \|q - w\|\}.
 \end{equation*}
 It is a set of cells, the union of which make up $Q$, where each point in cell $V_{\Gamma_{ij}}$ is closer to agent $i$ and agent $j$ than to any other pair of agents.
Similar to the work done by Jiang \textit{et al.} \cite{jiang2015higher} we will show, under certain conditions, that the second-order Voronoi partition is a minimising partition for $H_f$, i.e., $H_f(P, V(P)^\textbf{2}) \leq H_f(P, W)$ for any partition $W$. Consider the three following conditions on a general sensor model $f(x,y)$:
\begin{enumerate}
    \item[(C1)] $f(x+k, y) - f(x, y) \geq 0 \quad for \quad k \geq 0$
    \item[(C2)] $f(x, y+k) - f(x, y) \geq 0 \quad for \quad k \geq 0$
    \item[(C3)] $f(x,y) = f(y,x)$.
\end{enumerate}
Intuitively, (C1) and (C2) state that sensing performance should not improve as the distance from either agent increases, while (C3) states that the sensing abilities of all agents is homogeneous.
Condition (C3) is required for a well defined mapping between the unordered set $\Gamma_{ij}$ and the function $f(x,y)$. We will show that under these conditions the second-order Voronoi partition is a minimising partition for $H_f$, and then that the photogrammetry sensor model (\ref{eq:photogrammetrySensorFunction}) satisfies these conditions.

\textit{Proposition 1:} Under conditions (C1)-(C3), for all $q \in V_{\Gamma_{ij}}$ and $(i,j),(k,l) \in C$ then
\begin{equation*}
    f(\|q-p_i\|, \|q-p_j\|) \leq f(\|q-p_k\|, \|q-p_l\|).
\end{equation*}

\textit{Proof:} Within $V_{\Gamma_{ij}}$, $\|q-p_i\| \leq \|q-p_k\|$ and $\|q-p_j\| \leq \|q-p_l\|$ by definition, therefore from (C1) and (C2) then $f(\|q-p_i\|, \|q-p_j\|) \leq f(\|q-p_k\|, \|q-p_j\|)$ and $f(\|q-p_k\|, \|q-p_j\|) \leq f(\|q-p_k\|, \|q-p_l\|)$. Therefore, combining inequalities, the result follows. $\hfill\blacksquare$

\textit{Lemma 1:} For a sensor model satisfying (C1)-(C3), an arbitrary partition $W = \{W_{\Gamma_{12}}, W_{\Gamma_{13}},...,W_{\Gamma_{(n-1)(n)}}\}$ of $Q$, and an agent configuration $P$ then
\begin{equation*}
    H_f(P, V(P)^\textbf{2}) \leq H_f(P, W).
\end{equation*}

\textit{Proof:} Consider the set $V_{\Gamma_{ij}} \cap W_{\Gamma_{kl}}$ for $(i,j),(k,l) \in C$. For all $q \in V_{\Gamma_{ij}} \cap W_{\Gamma_{kl}}$ there holds $f(\|q-p_i\|, \|q-p_j\|) \leq f(\|q-p_k\|, \|q-p_l\|)$ by Prop. 1. 
Therefore integrating over this set for all $(i,j)$ yields the following inequality
\begin{equation*}
    \begin{aligned}
    \sum_{\forall (i,j) \in C} \int_{V_{\Gamma_{ij}} \cap W_{\Gamma_{kl}}} f(\|q-p_i\|, \|q-p_j\|) \phi(q)dq \leq \\ \sum_{\forall (i,j) \in C} \int_{V_{\Gamma_{ij}} \cap W_{\Gamma_{kl}}} f(\|q-p_k\|, \|q-p_l\|) \phi(q) dq.
    \end{aligned}
\end{equation*}
As $(\sum_{\forall (i,j) \in C} V_{\Gamma_{ij}} \cap W_{\Gamma_{kl}}) \subset W_{\Gamma_{kl}}$ then
\begin{equation*}
    \begin{aligned}
    \sum_{\forall (i,j) \in C} \int_{V_{\Gamma_{ij}} \cap W_{\Gamma_{kl}}} f(\|q-p_i\|, \|q-p_j\|) \phi(q)dq \leq \\  \int_{W_{\Gamma_{kl}}} f(\|q-p_k\|, \|q-p_l\|) \phi(q) dq.
    \end{aligned}
\end{equation*}
Applying the sum over all $(k,l)$ to both sides and observing that 
\begin{equation*}
    \sum_{\forall (k,l) \in C} \sum_{\forall (i,j) \in C} \int_{V_{\Gamma_{ij}} \cap W_{\Gamma_{kl}}} \cdot \quad dq = \sum_{\forall (i,j) \in C} \int_{V_{\Gamma_{ij}}} \cdot \quad dq,
\end{equation*}
we show the following inequality, proving the lemma
\begin{equation*}
    \begin{aligned}
    \sum_{\forall (i,j) \in C} \int_{V_{\Gamma_{ij}}} f(\|q-p_i\|, \|q-p_j\|) \phi(q)dq \leq \\ \sum_{\forall (k,l) \in C} \int_{W_{\Gamma_{kl}}} f(\|q-p_k\|, \|q-p_l\|) \phi(q) dq. \quad \hfill\blacksquare
    \end{aligned} 
\end{equation*} 

Finally we show that the photogrammetry sensor model in (\ref{eq:photogrammetrySensorFunction}) satisfies (C1)-(C3).

\textit{Proposition 2:} The function $f(x,y) = h(x,y)$, as defined in (\ref{eq:photogrammetrySensorFunction}), satisfies (C1)-(C3).

\textit{Proof:} Beginning with (C1), when $\|\begin{bmatrix} x & y \end{bmatrix}\|_\infty > r$, $f(x+k,y) - f(x, y) = 2(\mbox{diam}Q)^2 - 2(\mbox{diam}Q)^2 = 0$. When $\|\begin{bmatrix} x & y \end{bmatrix}\|_\infty \leq r < x+k$, 
$f(x+k,y) - f(x, y) = 2(\mbox{diam}Q)^2 - (x^2 + y^2) \geq 0$. When, $\|\begin{bmatrix}x & y & x+k \end{bmatrix}\|_\infty \leq r$, 
$f(x+k,y) - f(x, y) = 2xk + k^2 \geq 0$. Condition (C1) is therefore satisfied. Condition (C2) follows from analogous reasoning and (C3) is trivial. $\hfill\blacksquare$

Using the above results the task of minimising the photogrammetry cost function is equivalent to minimising the less general cost function in terms of the second-order Voronoi partition of Q
\begin{equation}
\label{eq:simplerCost}
    \min_{W} H_h(P,W) = H_h(P,V(P)^\textbf{2}).
\end{equation}
This will be useful later in showing that $H_h(P,V(P)^\textbf{2})$ is bounded by a cost function for which well formed controllers exist. 

\subsection{Controller Design}
A natural choice of agent controller to minimise (\ref{eq:simplerCost}) is
\begin{equation}
\label{eq:badController}
    u_i = -k\frac{\partial H_h}{\partial p_i},
\end{equation}
for $k>0$, where each agent follows a local path that globally reduces the cost. The cost $H_h$ may have stationary points of inflection for a range of $P$, as the sensor model in (\ref{eq:photogrammetrySensorFunction}) uniformly penalises points not overlapped by a pair of agents. 
For example, if a small perturbation in an isolated agent's position does not cause its FOV to overlap with another agent, the agent will have zero control input $\frac{\partial H_h}{\partial p_i} = 0$.
The controller (\ref{eq:badController}) hence does not necessarily drive $H_h$ to a local minimum.
Previous coverage work considers sensor models with \textit{strictly increasing} penalties, resulting in a cost function with minimizers that have large basins of attraction. 
For this reason we consider an auxiliary cost function with the sensor model $f(x,y) = g(x,y) = x^2 + y^2$
\begin{equation*}
    H_{g}(P, W) = \sum_{\forall W_{\Gamma_{ij}} \in W} \int_{W_{\Gamma_{ij}}} g(\|q-p_i\|, \|q-p_j\|)\phi(q)dq.
\end{equation*}
Jiang \textit{et al.} \cite{jiang2015higher} show that the optimal partition for this cost is the second-order Voronoi partition $V(P)^\textbf{2}$, and they provide a distributed controller that drive the agents to a local minimum configuration for $H_g$
\begin{equation}
\label{eq:controllers}
    u_i = -k\frac{\partial H_g}{\partial p_i} =  -k(p_i-\overline{C}_{{W}_i}),
\end{equation}
where $k>0$ and $\overline{C}_{W_i}$ is the additive centroid
\begin{equation*}
    \overline{C}_{W_i} = \frac{\sum\limits_{\{\forall \Gamma_{ij} \in V(P)^\textbf{2} | p_i \in \Gamma_{ij}\}}C_{V_{\Gamma_{ij}}}M_{V_{\Gamma_{ij}}}}{\sum\limits_{\{\forall \Gamma_{ij} \in V(P)^\textbf{2}| p_i \in \Gamma_{ij}\}}M_{V_{\Gamma_{ij}}}}
    .
\end{equation*}
The variables $C_{V_{\Gamma_{ij}}}$ and $M_{V_{\Gamma_{ij}}}$ are the mass and centroid of sub-region $V_{\Gamma_{ij}}$ respectively,
 \begin{equation*}
     M_{V_{\Gamma_{ij}}} = \int_{V_{\Gamma_{ij}}}\phi(q)dq
     \quad
     \quad
     C_{V_{\Gamma_{ij}}} = \frac{1}{M_{V_{\Gamma_{ij}}}}\int_{V_{\Gamma_{ij}}}q\phi(q)dq.
 \end{equation*}
This controller is distributed in terms of the second-order Voronoi partition, as each agent can calculate their own control input provided they can calculate their second-order Voronoi cells.
To calculate a Voronoi cell an agent must know the position of the additional agent assigned to the cell, and agents assigned to adjacent cells.

We show for agent configuration $P$ and partition $W$ that $H_{g}$ bounds the photogrammetry cost function $H_h$. As the second order Voronoi partition $V(P)^\textbf{2}$ is a minimizer for both costs, this allows the use of controller (\ref{eq:controllers}) to converge to an agent configuration $P$ that approximately minimise $H_h$. 

\textit{Lemma 2:} For $r \in [0, \mbox{diam}Q)$ and $\beta = \frac{r}{\sqrt{2}\mbox{diam}Q}$ then for agent configuration $P$ and partition $W$,
\begin{equation*}
    H_{g} \leq H_h < \frac{1}{\beta^2}H_{g}.
\end{equation*}

\textit{Proof: }As the cost functions differ only in sensor model we need $g(x,y) \leq h(x,y) < \frac{1}{\beta^2}g(x,y)$ for all $q \in W_{\Gamma_{ij}}$, where $x = \|q-p_i\|$ and $y = \|q-p_j\|$.
For brevity we define $z = \|\begin{bmatrix} x & y \end{bmatrix}\|_\infty$. Starting with the lower bound, when $z \leq r$, $h(x,y) = g(x,y)$. When $z > r$, $h(x,y) = 2(\mbox{diam}Q)^2 \geq (x^2 + y^2) = g(x,y)$, as $x,y \in [0, \mbox{diam}Q]$. Thus the lower bound follows.
For the upper bound, when $z \leq r$, $h(x,y) = (x^2 + y^2) < \frac{1}{\beta^2}(x^2 + y^2) = \frac{1}{\beta^2}g(x,y)$, as $\beta < 1$. When $z > r$, $h(x,y) = 2(\mbox{diam}Q)^2$ while 
\begin{equation*}
    \frac{1}{\beta^2}g(x,y) = 2(\mbox{diam}Q)^2 \frac{x^2 + y^2}{r^2} = 2(\mbox{diam}Q)^2 \frac{\|\begin{bmatrix} x & y \end{bmatrix}\|^2}{r^2}.
\end{equation*}
 As $\|\begin{bmatrix} x & y \end{bmatrix}\| \geq \|\begin{bmatrix} x & y \end{bmatrix}\|_\infty$ and $\|\begin{bmatrix} x & y \end{bmatrix}\|_\infty > r$, $h(x,y) < \frac{1}{\beta^2}g(x,y)$ and the upper bound is proved.
$\hfill\blacksquare$

\textit{Remark 1:} The scale factor $\beta$ gives an intuition on when minimising $H_{g}$ using (\ref{eq:controllers}) will approximately minimise $H_h$. As the radius $r$ of the agent's FOV increases relative to the size of the region $\mbox{diam}Q$, $\beta$ increases and the bound gets tighter.

\textit{Remark 2:} This controller does not enforce inter-agent collision avoidance. As mentioned in Section \ref{problem_spec}, a secondary safety controller can modify control inputs for safety. Alternatively, for large-scale problems, a height offset could be applied to each agent to prevent collisions, with negligible effect on an agent's FOV.

\section{EXPERIMENTS} \label{experiments}

This section gives preliminary results validating this approach.
Simulations demonstrate the agents converging to an imaging configuration over a density distribution, and validate the cost function bounds derived in the previous section. Photogrammetry experiments using configurations generated in simulation then validate the approach. Two real-world regions are constructed with groups of distinctive objects against relatively plain flooring (see Fig. \ref{fig:env_pics} for region 1). 
Functions of the form 
\begin{equation*}
    \phi_j(q) = \sum_{i=1}^{k_j} A_{ij} \exp{\frac{-(q-\mu_{ij})^T(q-\mu_{ij})}{2 \sigma_{ij}^2}},
\end{equation*}
where $j \in [1, 2]$, capture the feature density of the two regions, respectively. The variable $k_j$ is the number of object groups in each region, with $k_1 = 1$ and $k_2 = 3$, while $A_{ij}, \mu_{ij}$ and $\sigma_{ij}$ are the scalar amplitude, 2D center and scalar standard deviation, respectively, of a Gaussian peak that models the feature density of the objects in each scene. 
To construct these functions, images are captured at a range of locations in the region and the SIFT feature extraction algorithm is run on each. The feature count of each image is used as a measurement of the density at the point it was captured, and $\phi_j$ is constructed as a least squares fit, over the parameters $A_i, \mu_i$ and $\sigma_i$, to these measurements (see Fig. {\ref{fig:env_pics}).
The density functions are created and provided to the agents \textit{a priori}.

\begin{figure}
    \centering
    \smallskip
    \smallskip
    \subfigure[]{\includegraphics[width=0.49\columnwidth, height = 4cm]{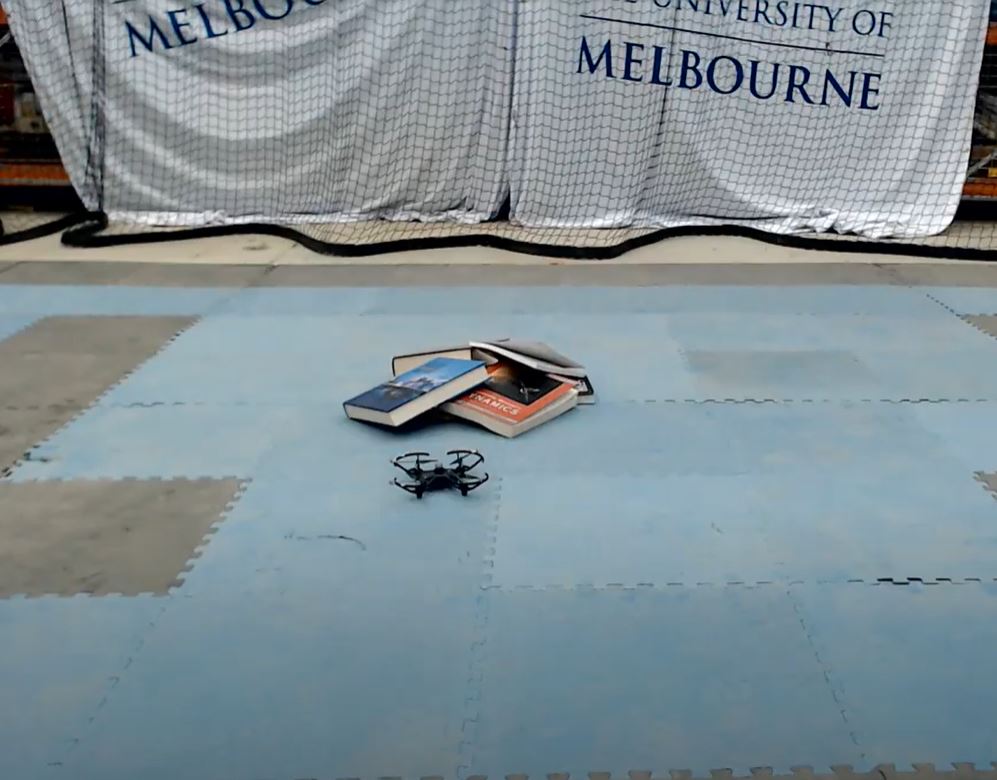}}
    \subfigure[]{\includegraphics[width=0.49\columnwidth, height=4cm]{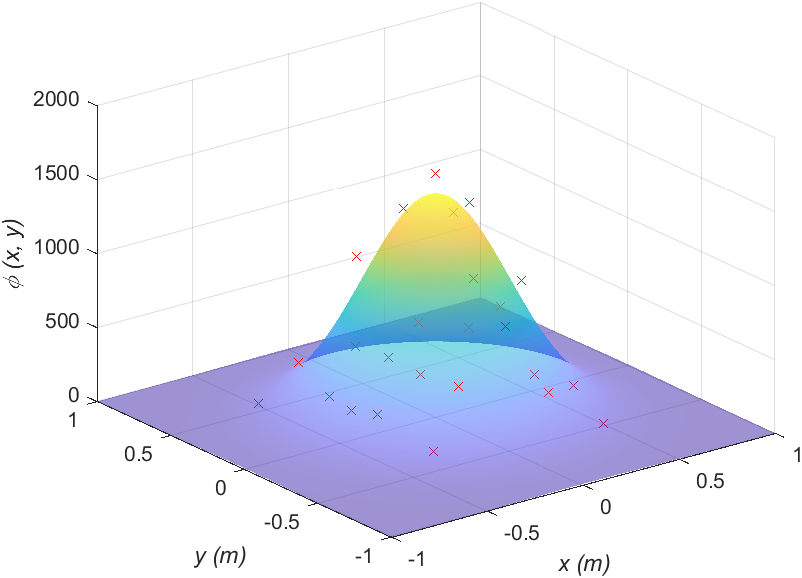}}
    \caption{(a) Real-world environment 1. (b) Density function $\phi_1$. The red crosses overlaying the density function are the measurements.}
    \label{fig:env_pics}
\end{figure}

\subsection{Simulations}

Agents are initialised randomly within $Q$ and converge over the density functions $\phi_1$ and $\phi_2$, using the controller in (\ref{eq:controllers}). For $\phi_1$, $n = 9$ and $16$ agents are used and for $\phi_2$, $n= 16$ and $20$ agents are used. In all simulations $r = 0.5$ and $\mbox{diam}Q = 1.5 \sqrt{2}$. As the agents approach their final configurations the auxiliary cost function $H_g$ is minimised and bounds the photogrammetry cost function $H_h$. Fig. \ref{fig:env_1_9_agents_plots} shows the agents trajectory, the second-order Voronoi partition and the density function, as well as the costs over time. It can be seen that $H_h$ is bounded by $H_g$ and $\frac{1}{\beta^2}H_g$, in agreement with Lemma 2, and decreases as $H_g$ is minimised. 
The photogrammetry cost $H_h$ for a uniform grid configuration of agents is plotted for comparison. The agents converge from a random configuration with higher cost than the grid configuration to an improved configuration with a lower cost.

\begin{figure}
    \centering
    \smallskip
    \smallskip
    \subfigure[$t = 0$s]{\includegraphics[width=0.68\columnwidth]{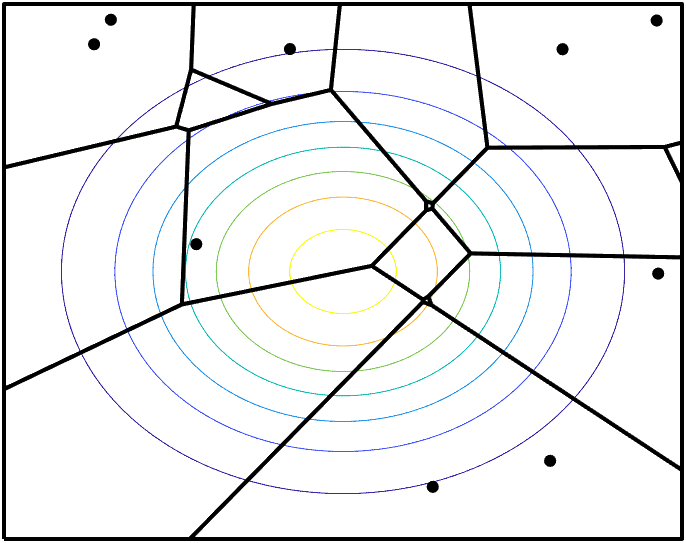}}
    \subfigure[$t = 5$s]{\includegraphics[width=0.68\columnwidth]{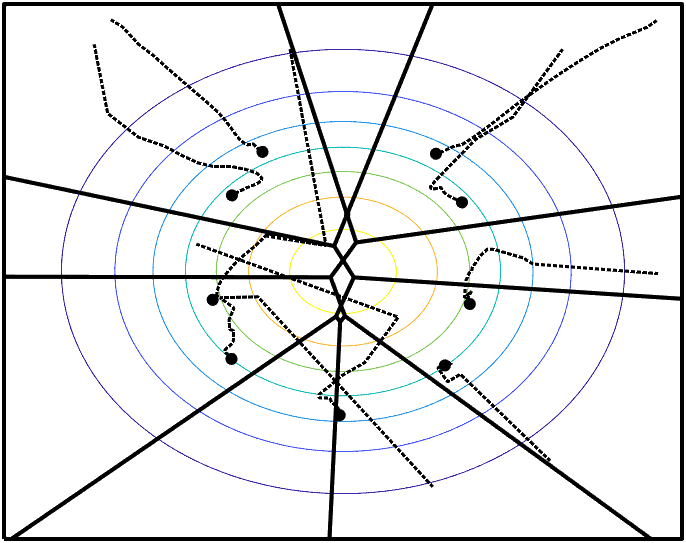}}
    \subfigure[Costs $H_h$ and $H_g$ over time]{\includegraphics[width=0.7\columnwidth]{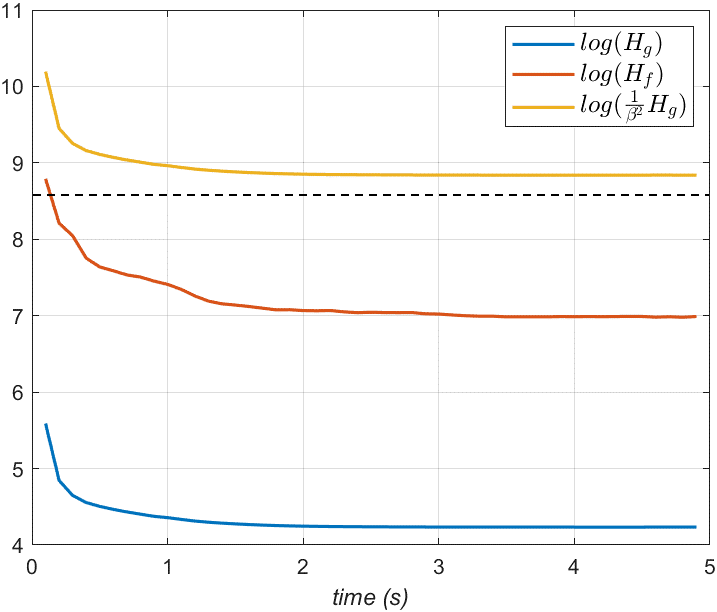}}
    \caption{MATLAB simulation for $n = 9$ and $\phi_1$. (a) Initial agent configuration (dots), second order Voronoi partition (lines), and density function (contours). (b) Agent trajectories (dashed) and final configuration (solid dots). 
    (c) Photogrammetry and auxiliary costs (solid lines), and cost of a grid configuration (dashed line).}
    \label{fig:env_1_9_agents_plots}
\end{figure}

\subsection{Photogrammetry Experiments}

Images of the real-world regions are captured from the final positions of the agents from simulations once (\ref{eq:controllers}) was close to converged. Image sets are also captured from a random configuration and a uniformly-spaced grid configuration. Each image set is used to generate a 3D point cloud model with photogrammetry. To compare the reconstruction quality the point count is used as a performance metric. Additionally, the number of images discarded by the photogrammetry pipeline due to insufficient feature overlap is reported to further analyse the configuration of each image set. Finally, textured meshes are generated from the point clouds to visually compare the performance and to emphasise the advantage of increasing the point count.

A DJI Tello EDU UAV is modified to have its camera facing downwards, and the Tello ROS Driver \cite{tello} is modified to allow for requesting and receiving high quality images. Velocity control is used to track the imaging points and localisation is achieved using a Vicon motion capture system. Once all images are gathered, the photogrammetry software Meshroom \cite{Meshroom} is used, with the standard pipeline applied, to generate a 3D reconstruction.

The results are given in Tables I-IV while Figures \ref{fig:env_1_16_agents_reconstructions} and \ref{fig:env_2_16_agents_reconstructions}  show the point clouds and meshes generated for $(\phi_1, n=16)$ and $(\phi_2, n=16)$ respectively. For all tests, the image sets generated via the coverage approach generated 3D models with a higher point count. 
For the larger environment $\phi_2$, the grid configuration did not give sufficient image overlap with 16 agents and nine images were discarded. With 20 agents the overlap was sufficient and zero images were discarded. In both cases the coverage configuration gave the highest point count, demonstrating how the approach adapts to the number of agents available.
Notably for $\phi_2$ and $n = 20$, despite more images being discarded in the coverage configuration than in the grid configuration, the coverage configuration yielded a higher point count. This demonstrates how the coverage configuration concentrated agents around feature-dense areas and produced a higher quality reconstruction from less images.

\begin{table}[!t]
\footnotesize
\centering
\smallskip
\smallskip
\caption{$\phi_1$, $n=9$ Results}
\label{tab:env_1_9_agents_results}
\begin{tabular}{|c|c|c|c|}
\hline
{} & \textbf{Random} & \textbf{Grid} & \textbf{Coverage} \\ \hline
\textit{Point Count} & 230 & 213 & 3140 \\ \hline
\textit{Images Discarded} & 6 & 6 & 0 \\ \hline
\end{tabular}

\centering
\caption{$\phi_1$, $n=16$ Results}
\label{tab:env_1_16_agents_results}
\begin{tabular}{|c|c|c|c|}
\hline
{} & \textbf{Random} & \textbf{Grid} & \textbf{Coverage} \\ \hline
\textit{Point Count} & 2967 & 2168 & 4140 \\ \hline
\textit{Images Discarded} & 3 & 3 & 0 \\ \hline
\end{tabular}

\centering
\caption{$\phi_2$, $n=16$ Results}
\label{tab:env_2_16_agents_results}
\begin{tabular}{|c|c|c|c|}
\hline
{} & \textbf{Random} & \textbf{Grid} & \textbf{Coverage} \\ \hline
\textit{Point Count} & 3511 & 894 & 3799 \\ \hline
\textit{Images Discarded} & 5 & 9 & 1 \\ \hline
\end{tabular}

\centering
\caption{$\phi_2$, $n=20$ Results}
\label{tab:env_2_16_agents_results}
\begin{tabular}{|c|c|c|c|}
\hline
{} & \textbf{Random} & \textbf{Grid} & \textbf{Coverage} \\ \hline
\textit{Point Count} & 2593 & 3903 & 4149 \\ \hline
\textit{Images Discarded} & 9 & 0 & 1 \\ \hline
\end{tabular}
\end{table}

\begin{figure}
    \centering
    \smallskip
    \smallskip
    \subfigure[]{\includegraphics[height=3.25cm,width=0.43\columnwidth]{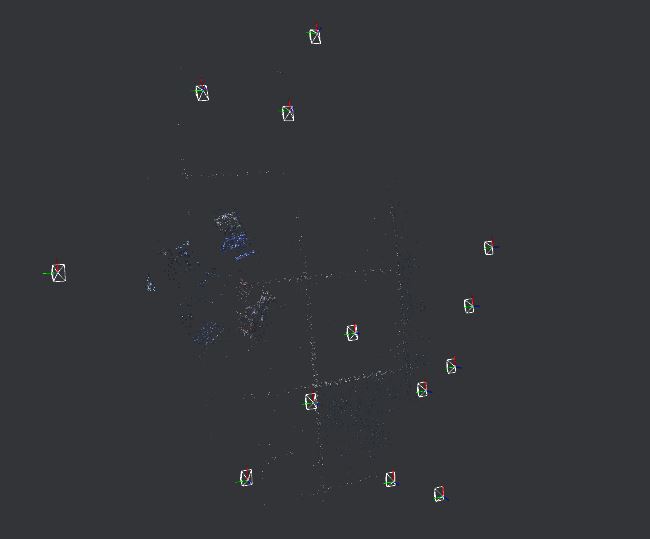}}
    \subfigure[]{\includegraphics[height=3.25cm,width=0.43\columnwidth]{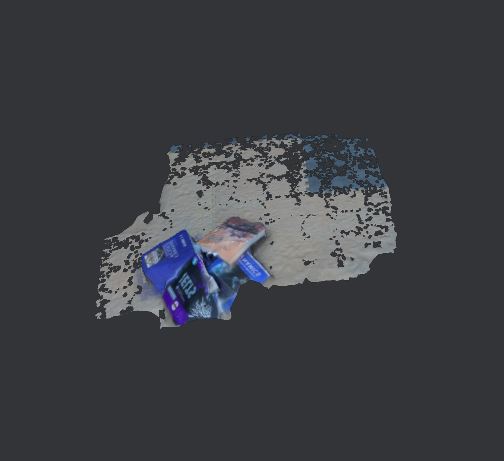}}
    \subfigure[]{\includegraphics[height=3.25cm,width=0.43\columnwidth]{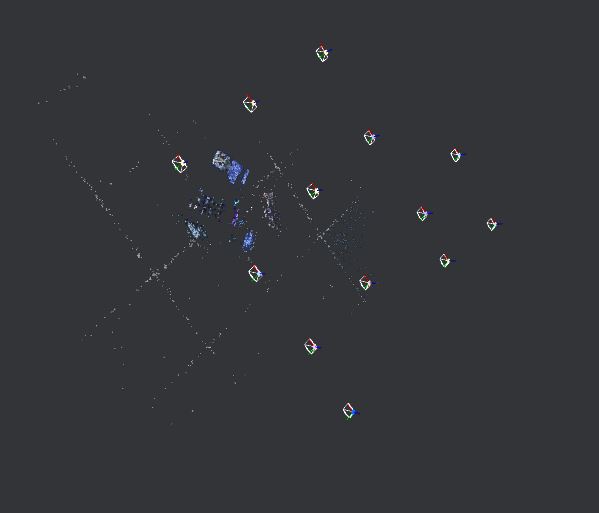}}
    \subfigure[]{\includegraphics[height=3.25cm,width=0.43\columnwidth]{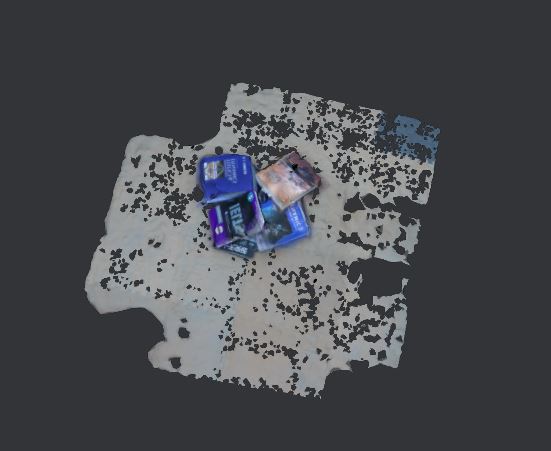}}
    \subfigure[]{\includegraphics[height=3.25cm,width=0.43\columnwidth]{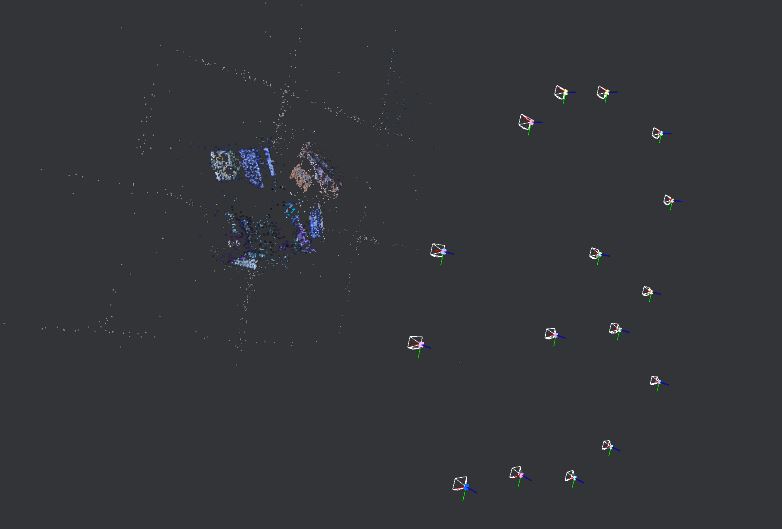}}
    \subfigure[]{\includegraphics[height=3.25cm,width=0.43\columnwidth]{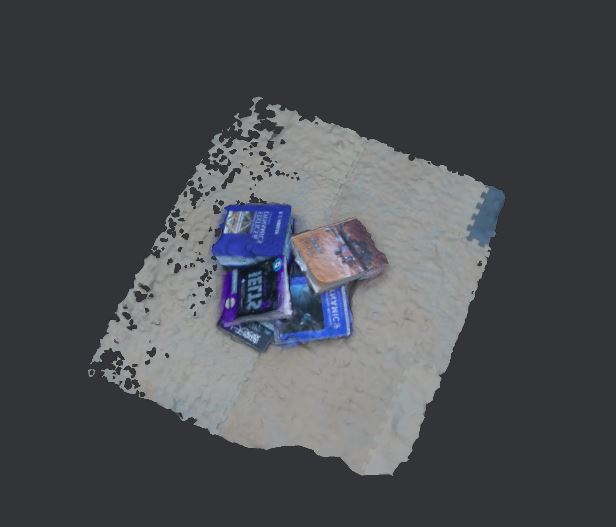}}
    \caption{Point clouds and textures for $\phi_1$ and $n=16$. (a-b) Random configuration, (c-d) grid configuration and (e-f) coverage configuration.}
    \label{fig:env_1_16_agents_reconstructions}
\end{figure}

\begin{figure}
    \centering
    \smallskip
    \smallskip
    \subfigure[]{\includegraphics[height=3.25cm,width=0.43\columnwidth]{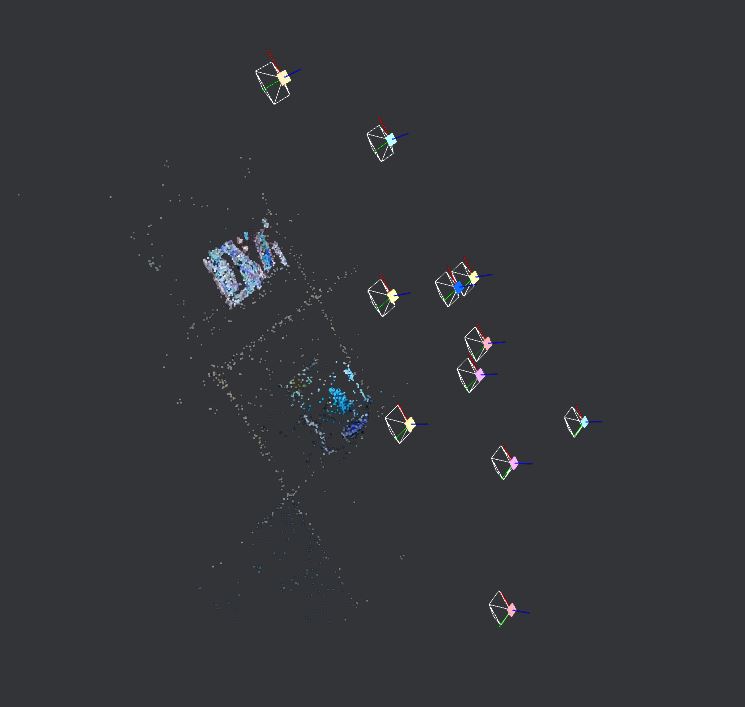}}
    \subfigure[]{\includegraphics[height=3.25cm,width=0.43\columnwidth]{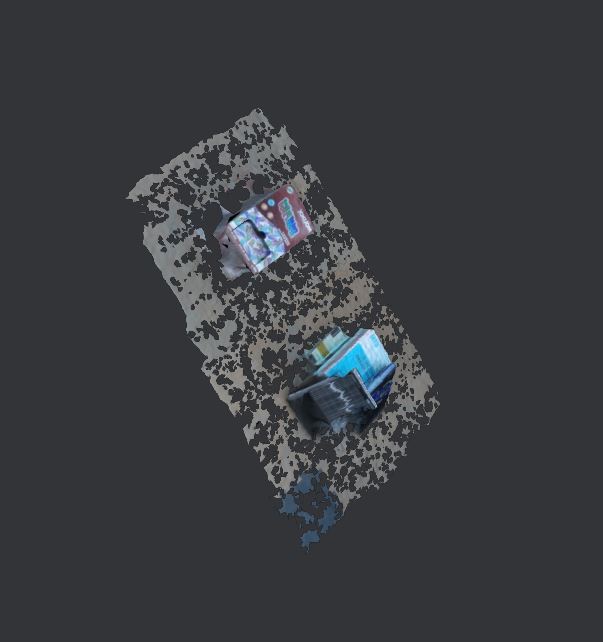}}
    \subfigure[]{\includegraphics[height=3.25cm,width=0.43\columnwidth]{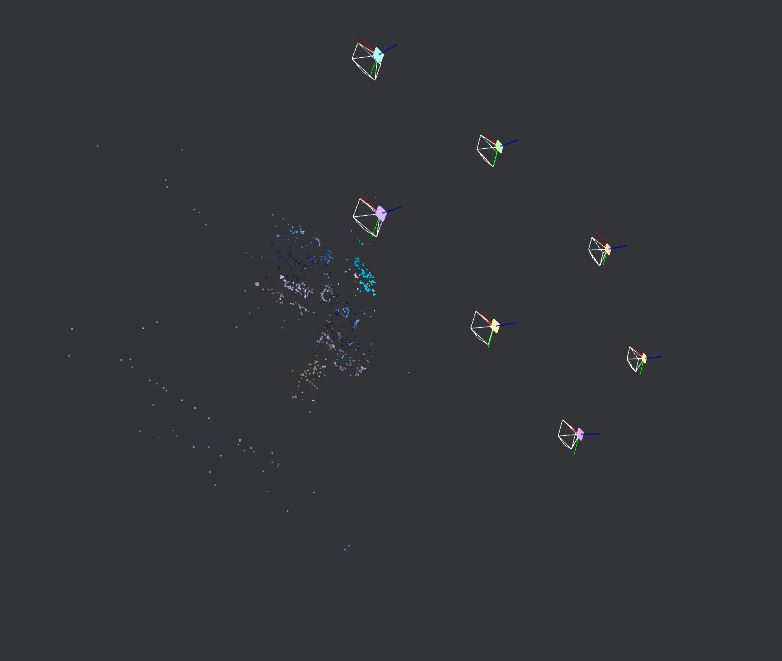}}
    \subfigure[]{\includegraphics[height=3.25cm,width=0.43\columnwidth]{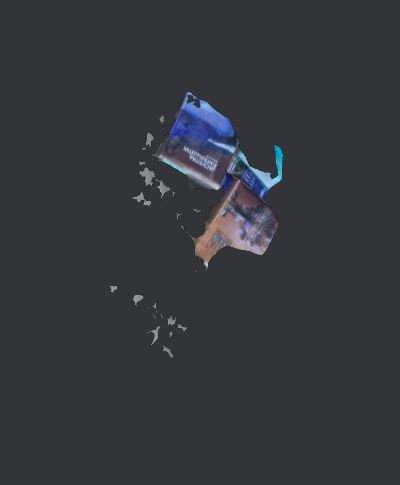}}
    \subfigure[]{\includegraphics[height=3.25cm,width=0.43\columnwidth]{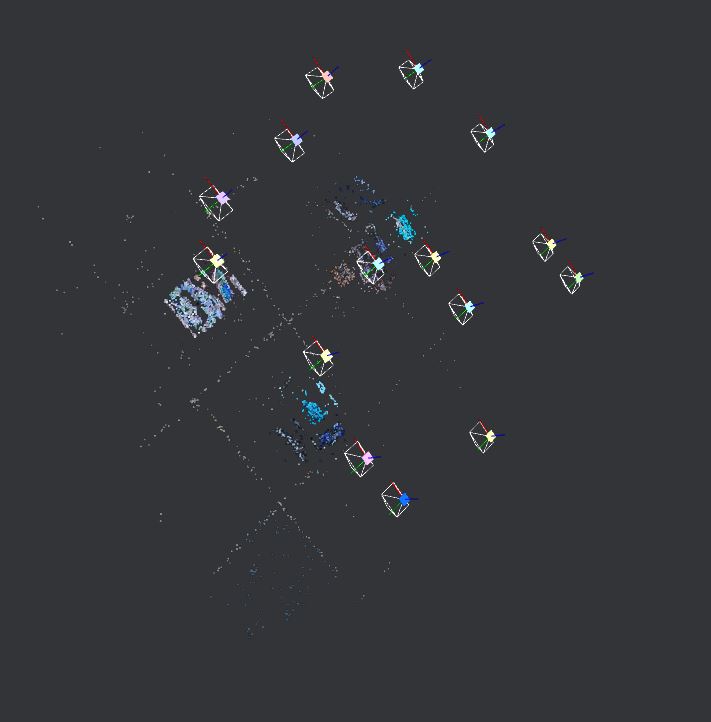}}
    \subfigure[]{\includegraphics[height=3.25cm,width=0.43\columnwidth]{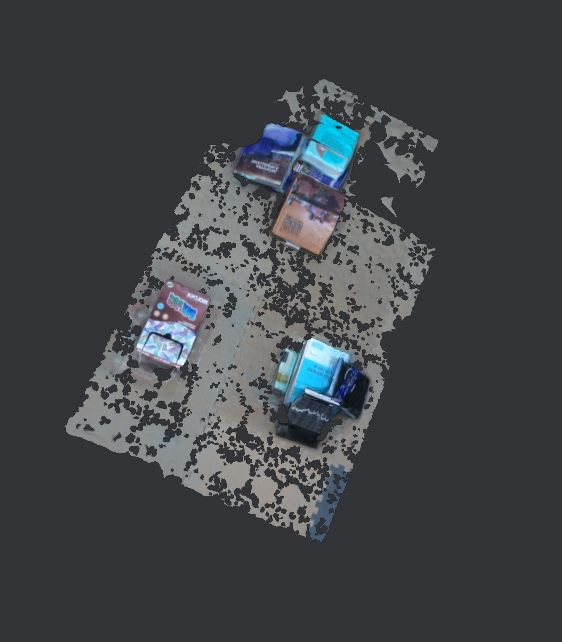}}
    \caption{Point clouds and textures for $\phi_2$ and $n=16$. (a-b) Random configuration, (c-d) grid configuration and (e-f) coverage configuration.}
    \label{fig:env_2_16_agents_reconstructions}
\end{figure}

\section{CONCLUSIONS}\label{conclusions}
This paper presents an approach to multi-agent UAV photogrammetry. A cost function for photogrammetry performance, inspired by the coverage control literature, is formulated. This cost includes a model of the real-world region as a feature-density distribution, and a sensor model capturing how distinctive real-world feature points must be captured in at least two images to be reconstructed. The photogrammetry cost is shown to be bounded by an existing cost function for which well formed controllers have been derived in the literature. We validate the bounded cost in simulation, and demonstrate the effectiveness of the method by generating photogrammetry reconstructions. 

Future work will explore more baseline photogrammetry methods and complex test instances for further validation of the approach.
Additionally, we plan to demonstrate this approach on a network of UAVs without prior knowledge of a density distribution, instead using run-time density estimation techniques \cite{schwager2009decentralized}. Finally, future work can incorporate each agent's altitude as a free variable, effecting the agent's FOV and image resolution.

\addtolength{\textheight}{-1cm}   

\bibliographystyle{IEEEtran}
\bibliography{bib}

\end{document}